# A New Approach to SMS Steganography using Mathematical Equations

Min Yang Lee, Vahab Iranmanesh*, Juan C. Quiroz
Faculty of Science and Technology
Sunway University
Bandar Sunway, Malaysia
12050431@imail.sunway.edu.my, vahab.iranmanesh@gmail.com, juanq@sunway.edu.my

*Abstract*—In the era of Information Technology, cyber-crime has always been a worrying issue for online users. Phishing, social engineering, and third party attacks have made people reluctant to share their personal information, even with trusted entities. Messages that are sent via Short Message Service (SMS) are easily copied and hacked by using special software. To enforce the security of sending messages through mobile phones, one solution is SMS steganography. SMS Steganography is a technique that hides a secret message in the SMS. We propose a new approach for SMS steganography that uses a mathematical equation as the stego media in order to transmit the data. With this approach, we can hide up to 35 characters (25%) of a secret message on a single SMS with maximum of 140 characters.

## I. INTRODUCTION

Steganography is a security method that has been implemented through generations. Steganography is a word stemming from the ancient Greek with the meaning of covered writing [1, 2, 4, 6, 7, 9, 10, 21]. The technique hides sensitive information into a cover media, such as a text file [1, 2, 4], an image [7, 8], a video [1, 7, 10] or even an audio file [7, 11, 16, 20]. The cover media is created to prevent a third party from becoming suspicious about hidden information being transmitted [17]. With stenography there are slight changes made to the cover media, which cannot be perceived by the human eye or ear if the embedding is well-designed [1, 2, 4, 7]. Fig. 1 shows the general framework of a Steganographic system. The main components of the Steganographic system are:

- Secret message (*s*): A message that required to be transmitted.
- Secret key (*k*): A key that required to embed the secret message (*s*).
- Cover media (*c*): A file or data that is used to embed the secret message (*s*).
- Stego media (*sm*): A file or data with the embedded secret message (*s*).

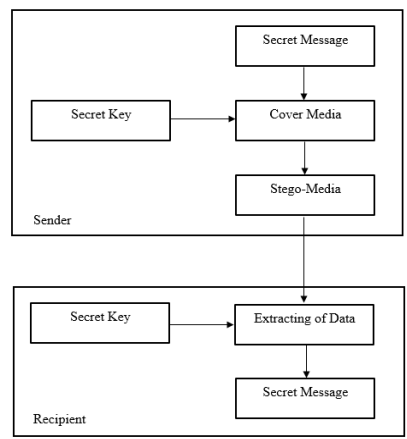

Figure 1. Steganographic System

Short message service (SMS) is one of the widely used services by mobile users [12, 15, 18, 19]. In particular, SMS steganography is classified as one of the most difficult hiding methods among other steganography techniques due to its less redundant bits [5]. In each SMS, only 140 characters can be sent, and most of the properties of the SMS texts are fixed. A mathematical equation is an equality containing one or more variables, but may also include mathematical symbols such as constants, operations, and functions. Equations can be simple or complex, and as long as expressions are well-formed, an equation may be short or as long as needed. In this paper, we propose a new approach in SMS Steganography using an equation the cover text to hide a secret message (*s*). Our approach can store up to 25% of the secret message (*s*) in each SMS.

The rest of this paper is organized as follows. Section 2 describes the related works in SMS steganography. Section 3 describes our proposed steganography method. Section 4 discusses our findings, and Section 5 presents conclusions and directions for future work.

## II. RELATED WORK

### A. Format Based Steganography

Format-based steganography utilizes the formatting of the cover media to hide a secret message (s). It modifies the existing format—such as white spaces, the size of the text, the printing type, etc.—to hide the secret message (*s*) [1, 2, 22]. This technique prevents suspicion from the naked human eye, but it cannot fool a computer. For instance, if the gap between two words contained two blank spaces, a computer can easily detect it. On the other hand, if the secret message (*s*) is hidden by resizing the text of the cover media, the computer may not be able to detect it, but it would be visible to the human eye. We next discuss several format-based steganography approaches.

*1) Line-Shifting*

In this method, the secret message (*s*) is hidden by shifting the line of the text vertically, up or down, or leaving the line unmoved [1, 3, 5, 11, 13]. The shifting of the line is not obvious to the human eye because it also shifts about 1/300 of an inch. This approach is suitable for printed text because the shifted line can be easily measured by using special measurement instruments. On the other hand, if the text is retyped or if Character Recognition Programs (OCR) are used on the cover text, the secret message (*s*) will be destroyed. Fig. 2 shows a fragment of the document using line shifting coding.

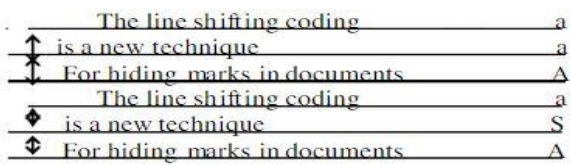

Figure 2. Line Shift Coding [1]

*2) Word-Shifting*

Word shifting entails hiding the secret message (*s*) between the gaps of words. In general, a single space is used between two words. In word shifting, we purposely insert two or more blank spaces to hide the secret message (*s*) [1, 3, 5, 11, 13]. This method is very common for printed text and is not obvious to the human eye. However, if someone carefully compared the modified text (stego text) and the original text (cover text), the difference would be obvious. Moreover, the information will also be destroyed by retyping or by using Character Recognition Programs (OCR). Fig. 3 shows a fragment of a document using word shifting.

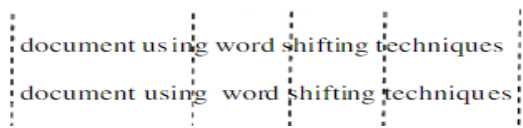

Figure 3. Word Shifting Example [1]

*3) Character Coding*

Character coding hides the secret message (*s*) by modifying features of the cover text. An example of feature coding is the end of the text character being sketched out or shortened in order to hide the secret message (s) [1, 5, 11, 13]. This method is able to store tremendous amount of information without raising suspicions. This is because people tend to think that the modified character is the result of a printing or formatting error. Fig. 4 shows a fragment of a document with character coding technique, where italics has been used to hide the secret message (s). Similar to the previous approaches discussed, the secret message (s) will be destroyed either by retyping or by using Character Recognition Programs (OCR).

```
There are more and less clever ways through which
information ca be concealed. Some of the cleverest are
those in which the fact that secret information is present
is not easily discovered. Of course, for such systems to be
genuinely useful, solutions yielded through sensible
decipherments must be clear and consistent.
```

Figure 4. Character Coding Example

*4) White Space*

This technique hides the secret message (*s*) using white space. The white spaces can be added either at the end of each sentence, at the end of a line, between words, or even after one paragraph of text [1, 5, 11, 13]. Having white spaces at the end of text is common, making it less suspicious to third parties. However, this method has a limitation of storing secret information. The hiding capacity is based on the amount of text in the file. If the text is scanned with a text editor with auto-correct, the white spaces can be easily detected thus destroying the secret message (*s*). Fig. 5 shows a fragment of the document using white space encoded text. Additional white space has been added to the end of each line to store the secret message (*s*).

```
T H E   Q U I C K   B R O W N   F O X
J U M P S   O V E R   T H E   L A Z Y
D O G .
         NORMAL
T H E   Q U I C K   B R O W N   F O X
J U M P S   O V E R   T H E   L A Z Y
D O G .
    WHITE SPACE ENCODED TEXT
```

Figure 5. White Space Encoded Text [1]

*5) Abbreviation*

This technique hides the secret message (*s*) using abbreviations. Abbreviations are the shortened form of a word or phrase. This approach is able to store little information into the text because the text are all shortened into few letters or even just single letter [3, 5, 11]. In practice, one must make sure that the used abbreviations are not arranged in a weird sequence that could gain suspicion. Table I shows examples of a few abbreviations.

*B. Linguistic Steganography*

Linguistic Steganography is a technique that modifies the character in the text instead of the formatting. Most often it will modify the size of character and the words to store the secret message (*s*) [1, 2]. In the following, several linguistic steganography approaches are discussed.

TABLE I. EXAMPLE OF ABBREVIATIONS

| Abbreviations | Meaning |
|---|---|
| R.S.V.P. | Respond, If You Please |
| P.S. | Post Script |
| E.T.A. | Estimated Time Of Arrival |
| B.Y.O.B. | Bring Your Own Bottle |

*1) Syntactic Method*

This method hides secret message ($s$) using punctuation marks, such as a period (.), a comma (,), a question mark (?), etc. [3, 5]. Using punctuation is a very common practice in SMS, but the challenge is making sure that the punctuation is used at the correct places to avoid suspicion. Fig. 6 shows a fragment of a document using syntactic technique to hide a secret message ($s$). The secret message ($s$) is hidden using the comma (,) and the question mark (?).

```
Good morning Albert how are you

Good morning Albert, how are you?
```

Figure 6. Syntactic technique example

*2) Word Spelling*

Word spelling technique explains how the secret message ($s$) is hidden in English words. The English spelling for some words in the United States (US) and the United Kingdom (UK) is different. For instance, 'Color' is the US spelling and 'Colour' is the UK spelling. In this case, we can use the additional 'u' to hide a secret message ($s$). This method is considered not secure because the US and the UK use some of these words differently [1, 3]. Hence, to avoid suspicions, only English words that look similar to both US and UK spelling can be used. Table II shows examples of the different spelling in US and UK for a few words.

TABLE II. AMERICAN VS BRITISH ENGLISH

| American English (United States) | British English (United Kingdom) |
|---|---|
| Center | Centre |
| Fiber | Fibre |
| Liter | Litre |
| Color | Colour |

C. *Text Steganography using Persian/Arabic Letters*

In this approach, the secret message ($s$) is hidden using Persian or Arabic letters. Persian and Arabic letters contain a number of points/dots in the letters, which allows hiding of the secret message ($s$) by altering the dots of the letters [1, 11, 13]. In the Persian language, there are a total of 18 letters out of 32 that have points/dots. Unfortunately, the approach is only applicable for Persian and Arabic users and not for English users. The English language only has 2 letters that include points/dots. So it is not ideal to implement it in English. Fig. 7 shows the alteration of a Persian/Arabic letter to store a secret message ($s$) by moving the dot slightly lower than the actual character.

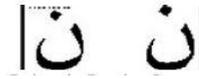

Figure 7. Persian/Arabic letter [1]

D. *SMS Steganography using Emoticons*

This approach hides the secret message ($s$) using emoticons, with each emoticon assigned to a single letter [5]. For instance, ':) :/ :(' can represent the secret message ($s$) 'you'. However, using emoticons in a roll to send a secret message ($s$) might gain suspicion from a third party, so this method is recommended to be used in multiple messages. Fig. 8 shows an example of using emoticons to map the secret message ($s$). The secret message ($s$) is "HELLO", with each emoticon representing one character.

```
Secret Message: HELLO

Emoticons: B-) =-O O:-) O:-) :-X
```

Figure 8. Example of using emoticons

E. *SMS Steganography using Sudoku Puzzle*

This technique hides a secret message ($s$) in a 9x9 Sudoku puzzle. The secret message ($s$) can be embedded into any rows and columns of the puzzle [14]. The recipient will then receive the Sudoku puzzle, and a special program is used to extract the secret message ($s$). The idea of storing the secret message ($s$) in the Sudoku puzzle is by permutations. The numbers will be the hint that stores the secret message ($s$). Fig. 9 shows a secret message ($s$) hidden in one of the rows or columns of the Sudoku puzzle.

```
Solve the Sudoku no.40

000:005:962
103:070:040
500:082:700
```

Figure 9: Sudoku technique example

III. PROPOSED METHOD

An equation is the most fundamental math statement. It uses equality to express the relation of values and mathematical operations. An equation can be formed with an arbitrary number of variables and mathematical symbols, such as constants, operations, and functions. Equations are frequently used in math related questions. Based on the question, the equations can be simple or complex. Therefore, we propose to use equations to hide a message communicated over SMS. By designing cover text ($c$) to look like a game or a math tutoring session, secret message ($s$) s can be hidden in equations.

Fig. 10 shows the proposed methodology framework using equations for SMS steganography. The next subsections explain the mapping, embedding, and extraction steps.

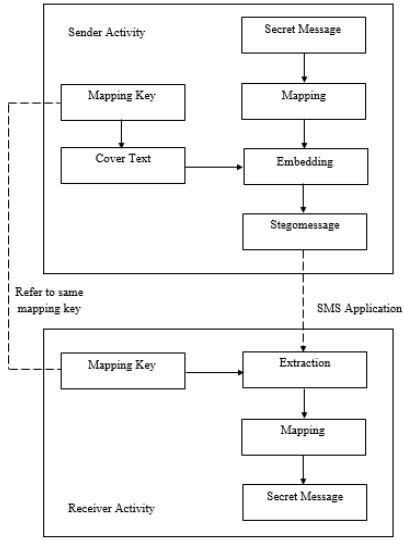

Figure 10. Framework of proposed idea

## A. Mapping

Our mapping technique is implemented by first changing the secret message ($s$) into random values, to further complicate message identification by a third party. Each letter of the English alphabet—uppercase and lowercase, the numbers 0-9, and whitespace, have their own values that are pre-set. Table III shows each uppercase and lowercase letter of the English alphabet mapped to a value between 1 and 52. Table IV shows the mapping of the numbers 0-9 and whitespace to the numbers 53-63. Finally, Table V shows the mapping of several operators to numbers. For instance, the secret message (s) "Attack now" would mapped to the values 1, 46, 46, 27, 29, 37, 63, 40, 41 and 49. Tables III-V are provided as example mappings, as each character may be mapped to different values in order to hide a message.

## B. Embedding

With the mapped values from the previous step, we next embed the values into an equation. In order to create the equation, we use a different key map ($k$). Table VI is an example of an operator key map ($k$), consisting of seven basic operators with a value that will change the mapped value when it is embedded in the equation.

The basic concept of hiding the mapped value is by randomly selecting operators, and adding the selected operators between the mapped values. For example, the mapped value from the previous step is 1, 46, 46, 27, 29, 37, 63, 14, 41 and 49, representing the secret message ($s$) "Attack Now". A first operator is randomly selected from Table VI, and added between the values 1 and 46. Then a second operator is randomly selected again from Table VI, and added between the values 46 and 46, and so on. After the last value, the last assigned operator will always be the equal operator in order to form an equation with standard formatting. An example of the resulting equation would be as in (1):

$$1\%46–46\wedge27\wedge29*37*63\wedge14*41*49= \quad (1)$$

We then convert equation (1) into a different equation by taking each number, and adding to the number the mapped value of the operator coming after the number. That is, the assigned operator value will be added to the value located at the front of each operator. For example, from Table V the mapped value for the modulo (%) operator is 62. The value 62 is then added to the number in front of the operator (the number 1), resulting in the number 63. Next, the mapped value of the subtraction operator, 5, is added to the number in front of the operator, 46, resulting in the value 51. This is repeated for all the operators, resulting in the final stego media (*sm*) shown in equation (2):

$$63\%51-220\wedge201\wedge107*115*237\wedge92*119*130= \quad (2)$$

Adding the mapped value of each operator to the mapped value of each letter further obfuscates the hidden message. Even if a third party were to obtain the mapping table for the letters and numbers, it would not be sufficient to break the hidden message. The third party would need to also obtain the operator key map in order to crack the secret message (s).

TABLE III. UPPERCASE AND LOWERCASE MAPPING TABLE

| Letter | Value | Letter | Value | Letter | Value | Letter | Value |
|---|---|---|---|---|---|---|---|
| A | 1 | N | 14 | a | 27 | n | 40 |
| B | 2 | O | 15 | b | 28 | o | 41 |
| C | 3 | P | 16 | c | 29 | p | 42 |
| D | 4 | Q | 17 | d | 30 | q | 43 |
| E | 5 | R | 18 | e | 31 | r | 44 |
| F | 6 | S | 19 | f | 32 | s | 45 |
| G | 7 | T | 20 | g | 33 | t | 46 |
| H | 8 | U | 21 | h | 34 | u | 47 |
| I | 9 | V | 22 | i | 35 | v | 48 |
| J | 10 | W | 23 | j | 36 | w | 49 |
| K | 11 | X | 24 | k | 37 | x | 50 |
| L | 12 | Y | 25 | l | 38 | y | 51 |
| M | 13 | Z | 26 | m | 39 | z | 52 |

The mapping table and the operator key map are examples. In practice, multiple mapping tables and key maps can be created and shared between the communicating parties. When the sender chooses one of the key maps, the key map selection is embedded into the pre-defined cover text as the number of points associated with a math quiz. For example, the following message hints to use key map 2: "Math Quiz (2 Pts) Answer:". With this hint, the receiver chooses the same key map in order to extract the secret message (*s*). Finally, the generated stego media (*sm*) will be the default cover text and the generated equation, such as: "Math Quiz (2 Pts) Answer: 63%51-220^201^107*115*237^92*119*130=".

TABLE IV. NUMBERS AND SPACE MAPPING TABLE

| Numeric | Mapping Value |
|---|---|
| 1 | 53 |
| 2 | 54 |
| 3 | 55 |
| 4 | 56 |
| 5 | 57 |
| 6 | 58 |
| 7 | 59 |
| 8 | 60 |
| 9 | 61 |
| 0 | 62 |
| " " | 63 |

TABLE V. OPERATOR KEY MAP

| Operators | Positive Value (Embedding) | Negative Value (Extracting) |
|---|---|---|
| ^ (Exponentiation) | 174 | -174 |
| + (Addition) | 32 | -32 |
| - (Subtraction) | 5 | -5 |
| * (Multiplication) | 78 | -78 |
| / (Division) | 100 | -100 |
| % (Modulo) | 62 | -62 |
| = (Equals) | 81 | -81 |

*C. Test Application*

Fig. 11 shows an Android application that was developed to test our SMS steganography method. The application includes two interfaces, one for sending a secret message ($s$) and another for retrieving a secret message (s). The sender inputs the secret message ($s$) on an input field and chooses the key map to embed the secret message ($s$). The application generates the stego media ($sm$) at the bottom of the interface, allowing the user to copy and send the stego media ($sm$) via any messaging application. Fig. 12 illustrates the interface for extracting the secret message ($s$) from the stego media ($sm$). The stego message ($s$) is copied from the corresponding messaging application and pasted into the input field. The number of points associated with the quiz is used to select the key map. The extracted secret message ($s$) is selected at the bottom of the interface.

IV. DISCUSSION

The idea of using an equation as part of the stego media ($sm$) can be challenging in terms of capacity, as each character of the secret message (s) ($s$) will be consuming about 3 to 4 characters in a 140 character SMS.

Table VI shows that the initial cover text—"Math Quiz (2 Pts) Answer: "—has 26 characters, which occupies 19% of the 140 characters of the SMS. The maximum length of the secret message ($s$) that can be hidden per SMS is 31 characters, which is about 22%. In addition, the capacity can be increased by reducing the length of the equation, but there will be a tradeoff between security and capacity. By shortening the characters representation for each secret message ($s$), the equation can hide more secret characters, but it will raise the level of suspicion.

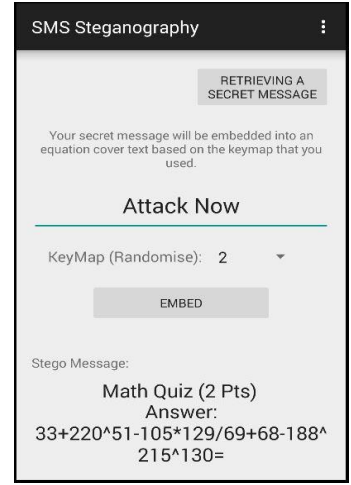

Figure 11. Generating a stego media based on an equation.

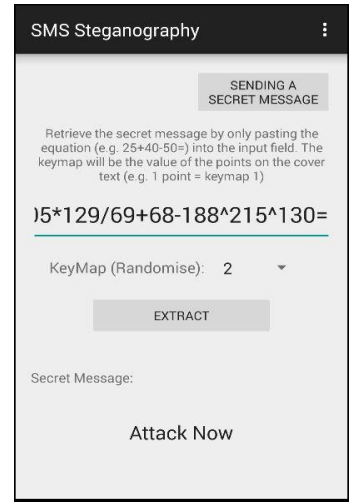

Figure 12. Extracting a hidden message from a stego media.

As compared to the format based steganography techniques, one of our advantages is that the secret message ($s$) will not be destroyed either by using OCR or retyping the stego media ($sm$). As compared to the system proposed in [5], where the sender has to create sentences with emoticons that both make sense and do not raise suspicion, the proposed system helps in creating the stego media ($sm$) that has a low level of suspicion to third parties and hackers.

TABLE VI. AMOUNT OF SECRET MESSAGE (S) AND CAPACITY USED

| Secret message (s) | Number of Characters | Cover Text Used (140 characters MAX) | Capacity Used (%) |
|---|---|---|---|
| (No message) | 0 | 26 | 19% |
| RUN | 3 | 39 | 27% |
| Sunway | 6 | 50 | 35% |
| High Five | 9 | 60 | 43% |
| I Love Sunway | 13 | 75 | 54% |
| Kill him ASAP after noon | 24 | 113 | 81% |
| Kill him ASAP after noon thanks | 31 | 139 | 99% |

Comparing to another proposed system, in [14] the capacity is limited because the secret message (*s*) is hidden in only one of the rows or columns of the 9x9 Sudoku puzzle. In our proposed system, the equation gets longer with the amount of secret message (*s*) being hidden without a size limitation, and it minimizes suspicion because equations come in varying sizes and values.

One of the limitations of our proposed system is that we make the assumption of one way communication. Although the stego message (*sm*) is sending a quiz to the receiver, we do not expect the receiver to send a response (an answer to the quiz) to the sender. From a hacker's point of view, the communication will seem as a math tutoring session where a teacher is broadcasting questions to a numbers of students.

Another limitation of the equation based SMS steganography is that if the equations are too random, this itself may raise suspicions. For example, it is unusual to have a math question where a number is raised to the power of 107 or 92, because the answer would be astronomical. Thus, it would be beneficial to limit the use of certain operators, make the selection of operators less random, or use single digit mapped values. All of these factors would make the equations more realistic, the corresponding answers would be reasonable, which in turn would make the communication less suspicious.

## V. CONCLUSIONS AND FUTURE WORK

This paper proposes a new SMS steganography approach using mathematical equations. In this method, the secret message (*s*) is embedded in the equation and combined with a pre-defined cover text to create the stego media (*sm*). With both of the combination of the pre-defined text and equation, hackers and other third parties would likely presume that the communication is a tutoring or a game session between two texters, without raising suspicions.

For future work, we plan to support more symbols and special characters. We would also like to address the use of Chinese or Persian characters in the secret message (s). For our test application, we intend to implement two way communication between sender and receiver. Finally, different types of equations, such as linear equations, polynomial equations, etc from other domains, such as physics equations and chemical equations, have the potential to provide a rich character set for hiding secret message (s) s (*s*).